\documentstyle [twoside,12pt] {article}

\newcommand{\bye}{\end{document}}

\newcommand{\baufg}{\begin{description}}
\newcommand{\eaufg}{\end{description}}
\newcommand{\bteilaufg}{\begin{description}}
\newcommand{\eteilaufg}{\end{description}}

\newcommand{\be}{\begin{equation}}

\newcommand{\ee}{\end{equation}}
\newcommand{\bes}{\begin{eqnarray}}
\newcommand{\ees}{\end{eqnarray}}
\newcommand{\ema}{\end {array} \right)}
\newcommand{\bma}{\left( \begin {array}}
\newcommand{\pslash}{\kern 0.2 em p\kern -0.45em /}
\newcommand{\dslash}{\kern 0.2 em \delta\kern -0.45em /}
\newcommand{\sla}[1]{\kern 0.2 em #1\kern -0.45em /}
\newcommand{\ra}{\rightarrow}
\newcommand{\ts}{\textstyle}
\newcommand{\scp}{\scriptstyle}

\newcommand{\rl}{I\!\!R}
\newcommand{\ler}{\stackrel{\scp <}{\scp\sim}}

\newcommand{\refe}[1]{(\ref{#1})}

\newcommand{\bt}{\begin{tabbing}
            \hskip 7.1 true cm \=\hskip 7.1 true cm \kill}
\newcommand{\et}{\end{tabbing}}
\newcommand{\bfig}{\begin{figure}}
\newcommand{\efig}{\end{figure}}

\begin{document}
\begin{titlepage}
\setcounter{page}{1}

\title{\vspace{0.2cm} $E_6$ GUT and Large Neutrino Mixing}
\author{Yoav ACHIMAN\\[0.2cm]
        Department of Physics \\
        University of Wuppertal \\
        Gau\ss{}str. 20, D-5600 Wuppertal 1\\
        Germany\\[0.8cm]
        Andr\'e LUKAS\\[0.2cm]
        Physic Department\\
        Technical University of Munich\\
        James-Franck-Str., D-8046 Garching near Munich\\
        Germany\\[0.1cm]
        and\\[0.1cm]
        Max-Planck-Institut f\"ur Physik\\
        Werner-Heisenberg-Institut\\
        P.~O.~Box 40 12 12, Munich\\
        Germany}

\date{August 1992}

\maketitle

\setlength{\unitlength}{1cm}
\begin{picture}(5,1)(-12.0,-19.0)
\put(0,0){WUB 92 - 20}
\put(0,-0.5){TUM - TH - 149/92}
\put(0,-1.0){MPI - Ph/92 - 69}
\end{picture}

\begin{abstract}
 All experimental results concerning possible neutrino oscillations
 are naturally and simultaneously accounted for in an $E_6$ GUT model. The
 fermionic mass matrices are dictated by the symmetry breaking and specific
 radiative corrections and not by the use of ``Ans\"atze'' or discrete
 symmetries.
\end{abstract}

\thispagestyle{empty}
\end{titlepage}
\clearpage

\setcounter{page}{1}

In a recent paper~\cite{paper1,old_papers} (hereafter~: ``paper I'')
we presented in detail an $E_6$ GUT with a very
specific set of mass matrices. In particular, the ``light'' neutrino mass
matrix is practically dictated by the quark sector and the scale of the
intermediate symmetry breaking. For an intermediate scale of
$10^{10} - 10^{12}$ GeV, suggested by the recent values of
$\sin^2\Theta_W$~\cite{weinberg_angel}, our ``solutions'' for the
neutrino masses and mixing were concentrated exactly around the value,
resulting from the latter announced GALLEX experiment~\cite{GALLEX}.
Moreover the experimental requirement of large neutrino
mixing allowed us to fix the favored breaking chain.

In this letter, we would like to emphasize this fact and use the exact GALLEX
results to limit the range of our solutions. This will enable us to fix
the hierarchy of the heavy VEV's and in particular the allowed values of the
intermediate breaking scale, which is also the scale of the right-handed (RH)
neutrinos.

Our model is based on the following considerations:
\begin{enumerate}
\item The superstrong  breaking of $E_6$ will be generated by one or
several symmetric $\Phi_{\bf 351}$. This dictates the direction of
the breaking, it must go via $SO(10)$, the only maximal
subgroup with singlets in those representations. The further breaking
goes then through $SU(5)$ or subgroups of
$G_{PS}=SU_C(4)\times SU_L(2)\times SU_R(2)$.

\item The low energy breaking into $SU_C(3)\times U_Q(1)$ will
be generated via one $H_{\bf 27}$ Higgs representation. In this case
all the mass matrices of the standard fermions are proportional to
each other on the tree level and can be diagonalized simultaneously
\be
 \begin{array}{lp{2cm}l}
  \hat{M}_e^0=\hat{M}^0&&\hat{M}_d^0=\hat{M}^0\\[0.5cm]
  \hat{M}^0_{\nu,Dir}=a\hat{M}^0&&\hat{M}_u^0=a\hat{M}^0
 \end{array}
 \label{tree_level_modell}
\ee
with
\bes
 \hat{M}^0&=&{\rm diag}(\mu_1,\mu_2,\mu_3) \label{m0_def}\quad \mu_i
 \in \rl .
\ees

\item The main requirement of our model is that the one-loop contributions
to the mass matrices are dominated by the diagrams
of fig.~\ref{fig : e6_rc1}. Those diagrams involve in the loop superheavy
gauge bosons and fermions but no scalars. This requirement can be justified
using arguments of maximal calculability and predictability. Contributions
involving scalar loops are less predictable than those involving gauge
bosons with known couplings. Only in supersymmetric theories with SUSY
broken at a low energy scale, the scalars may play an important role. In
those theories, however, the non--renormalization theorems allow one to
avoid unwanted couplings in the superpotential, to all orders. This
amounts then to fixing by hand the allowed radiative corrections.
Our requirement assumes therefore that SUSY is broken at a relatively
high scale and that all relevant Yukawa couplings are much smaller than
the gauge couplings~\footnote{This is well known phenomenologically
for the Yukawa couplings of standard fermions except for the top.}.\\
Calculability dictates at the same time one more important requirement.
It is well known that two-loop corrections to the diagram in
fig.~\ref{fig : e6_rc1}
diverge (see fig.~\ref{fig : e6_rc1_cl}). An obvious way to avoid this
\begin{figure}
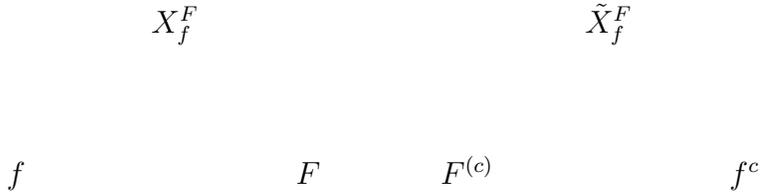

 \begin{center}
 \begin{tabular}{p{1.5cm}p{1.5cm}p{1.5cm}p{1.5cm}p{1.5cm}p{2cm}}
  &$v_{F,f}^0\quad =$&$O(M_W)$&$v_{F,f}\quad =$&$O(M_X)$&\\[1.5cm]
  &$X_f^F$&&&$\tilde{X}_f^F$&\\[1.5cm]
  $f$&&$F$&$F^{(c)}$&&$f^c$\\
 \end{tabular}
 \end{center}
 \caption{One-loop corrections to standard fermion masses containing
          gauge bosons.}
 \label{fig : e6_rc1}
\end{figure}
\begin{figure}
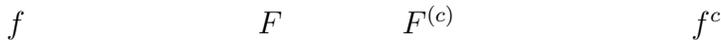

 \vspace{3cm}
 \begin{center}
 \begin{tabular}{p{1cm}p{1.5cm}p{1.5cm}p{1.5cm}p{1.5cm}p{1cm}}
  $f$&&$F$&$F^{(c)}$&&$f^c$\\
 \end{tabular}
 \end{center}
 \vspace{2cm}
 \caption{Infinite second order contributions to Higgs self coupling.}
 \label{fig : e6_rc1_cl}
\end{figure}
problem~\cite{old_papers} is to require that the superheavy mass terms are
``orthogonal'' to the tree level masses of the light fermions, in the
family space. In other words, one obtains a calculable theory if in the
framework of diagonal tree level mass matrices the radiative contributions
are pure off-diagonal. The off-diagonal corrections induced by the
diagrams of fig.~\ref{fig : e6_rc1} are given by~\cite{ibanez}
\bes
 \delta M_{F,f}&=&w_{F,f}M_F \nonumber\\
 w_{F,f}&=&\frac{3\alpha}{4\pi}\frac{k_{F,f}\alpha_g^{F,f}\beta_g^{F,f}}
           {m_{F,f}^2-\tilde{m}_{F,f}^2}\ln \left( \frac{m_{F,f}}
           {\tilde{m}_{F,f}} \right)^2. \label{corr_type1}
\ees
$M_F$ is the mass matrix of the superheavy fermion $F$, $m_{F,f}$ and
$\tilde{m}_{F,f}$ are the masses of the gauge bosons $X_f^{F,\pm}$ and
$\tilde{X}_f^{F,\pm}$. $k_{F,f}$ is the gauge boson mixing factor and
$\alpha_g^{F,f}$, $\beta_g^{F,f}$ are the group theoretical coefficients
of the gauge couplings. Since one of the gauge bosons carries a weak
isospin $I_W=1/2$, one VEV contributing to the mixing $k_{F,f}$ is of
order $M_W$~:
\be
 \begin{array}{ll}
  k_{F,f}=O(v_{F,f}^0v_{F,f})&\;v_{F,f}^0=O(M_W).
 \end{array}
 \label{gauge_mixing}
\ee
The corrections are in general of the order of magnitude $\alpha M_W$,
depending on the ratios of the superheavy masses in eq.~\refe{corr_type1}.
In paper~I it is proven that in our breaking scheme  $F=N,\nu^c$ give the
leading off-diagonal contributions\footnote{This is also true in the case
of a superheavy $SO(10)$--invariant VEV in the
${\bf 27}_{E_6}$--Higgs--representation. At the first sight this looks
dangerous because such a VEV generates large masses for the fermions in
${\bf 10}_{SO(10)}$~\cite{barb_nan1}. But for group theoretical
reasons contributions
from these fermions are restricted to the combinations $(e,D)$ and
$(d,E)$ (where $D$ and $E$ are the charged exotic particles in the
${\bf 10}_{SO(10)}$). In both cases the large VEV $v$ responsible for
gauge boson mixing breaks $SU(5)$ and consequently leads to a suppression
by the large $SO(10)$ invariant masses of the corresponding gauge bosons
(see eq.~\refe{corr_type1}).} where $N$ is the $SO(10)$--singlet in
${\bf 27}_{E_6}$. For group-theoretical reasons the graphs
with $F=N$ contribute to the masses of all standard fermions, whereas those
induced by $F=\nu^c$ are limited to the $u$-mass matrix~\cite{barb_nan1}.
Taking these radiative corrections into account we get the following
mass matrices :
\be
 \begin{array}{lp{2cm}l}
  M_e=\frac{\ts 1}{\ts r}(\hat{M}^0+E)&&M_d=\hat{M}^0+pE\\[0.5cm]
  M_{\nu,Dir}=\frac{\ts 1}{\ts r}(a\hat{M}^0+sE)&&M_u=a\hat{M}^0+qE+\Delta\;.
 \end{array}
 \label{model}
\ee
The matrices $E$ and $\Delta$ represent the contributions induced by $F=N$
and $F=\nu^c$ respectively, while their relative strengths are given by
$p,q$ and $s$. The different renormalization behaviour of the quarks
and leptons is taken into consideration by the factor\footnote{For large
top masses this is not a very good approximation, see remark about
that latter.} $1/r$.\\

\item The matrix $\Delta$ is clearly proportional to the RH-neutrino mass
matrix
$M_{\nu,R}$~:
\bes
 M_{\nu,R}&=&\eta \Delta\\ \label{r_neutr_mass}
 \eta&=&\frac{1}{w_{\nu_c,u}}\; . \label{eta_def}
\ees
$\eta$ is very large and the see saw mechanism~\cite{see_saw} is
naturally realized. The mass matrix of the light neutrinos is therefore
\be
 M_{\nu,small}\simeq\frac{1}{\eta}M_{\nu,Dir}\ \Delta^{-1}\ M_{\nu,Dir}.
 \label{nu_small}
\ee
Now, because $N$ and $\nu^c$ are Majorana particles the correction
matrices must be symmetrical. They are off-diagonal and in general complex.
\end{enumerate}

We analyzed first the $M_u, M_d, M_e$ matrices of eq.~\refe{model}
for real entries. In this case, as explained in paper~I,
there is only one ``free'' parameter
that obeys certain theoretical restrictions. We looked for ``solutions''
which give the best $\chi^2$ fit to the known masses of the quarks and the
charged leptons~\cite{GL} as well as the CKM matrix~\cite{GN}. We
found solutions which allow top masses up to $m_t \sim 250$ GeV. This
result is not trivial because it is very sensitive to the set
of experimental input masses used. Our results were obtained using
the by now standard set of masses due to Gasser and Leutwyler~\cite{GL}
while only a slightly different set of masses due to Barducci
et al.~\cite{gatto} allowed for $m_{t,phys}\ler 50$ GeV only.

We showed then in paper I that giving $\Delta$ a phase,
one can have CP violation without changing essentially the above results.

By studying different breaking chains of $E_6$ we can limit the freedom in the
solutions drastically.  We considered different possibilities,
in particular~:
\begin{itemize}
 \item[a)] One independent correction matrix, i.e. only one $\Phi_{\bf 351}$
used or the superstrong breaking follows the chain
 $E_6\ra SO(10)\ra  G_1\ra ... \ra G_n \ra SU_C(3)\times SU_L(2)\times U(1)$
 where $G_i\subset G_{PS} = SU_C(4) \times SU_L(2) \times SU_R(2)$.
 \item[b)] Two independent corrections matrices, i.e. more than one
$\Phi_{\bf 351}$ used but the superstrong breaking going via
 $E_6\ra SO(10)\ra SU(5) \ra SU_C(3)\times SU_L(2)\times U(1)$.
\end{itemize}
The essential result of this study is that these two breaking chains
favor in our model top masses above 100 GeV !~~Case a)
is even more restricted~:
\[
105 \hbox { GeV}\ler m_{t,phys}\ler 125 \hbox { GeV}.
\]
The neutrino sector plays a special role in the model. Given the superstrong
breaking chain, including the corresponding hierarchy of the superheavy VEV's,
 the neutrino properties are fixed. This is related to the fact that the mass
matrix of the RH neutrinos is explicitely given, in
eq.~\refe{r_neutr_mass}, once the quark-sector is solved.
In view of the freedom in the quark-sector ($m_t$, phases of the CKM matrix,
etc.) we can predict only the order of magnitude of the neutrino masses. This
is the reason why we could use for the renormalization factors between leptons
and quarks one representative parameter $r$, which is fixed by self-consistency
arguments in paper I to be
$ 2.7 < |r| <  3.3 $. This means actually that for very large top masses the
value of $m_{\nu_{\tau}}$ may be slightly higher than our results. Using the
calculations of ref.~\cite{Bludman} we expect an additional factor of 2 or
so.\\
In paper I we got the following basic results in the neutrino sector~:
The chain a) i.~e.~$E_6\ra SO(10)\ra  G_{PS}\ra ...$ or the
usage of only one $\Phi_{\bf 351}$ representation give only
small $\nu$--mixing $\ler 10^{-3}$. Consequently one needs several
$\Phi_{\bf 351}$ representations and the alternative breaking chain b)~:
$E_6\ra SO(10)\ra SU(5) \ra SU_C(3)\times SU_L(2)\times U(1)$.
To have more predictability in last breaking chain we studied the two
special cases~:
\begin{itemize}
 \item[1)] $|H(1,1)|\ll|\lambda_i^2\phi^i(1,1)|$
 \item[2)]  $|H(1,1)|\gg|\lambda_i^2\phi^i(1,1)|$.
\end{itemize}
where we used the following definitions~: $H(1,1)$ and $\phi^i(1,1)$
denote the $SO(10)$-invariant VEV's of $H_{\bf 27}$ and
$\Phi^i_{\bf 351}$ respectively and $\lambda_i$ the effective coupling
constant for the mixing between $H_{\bf 27}$ and $\Phi^i_{\bf 351}$.
Then the parameters in eq.~\refe{model} receive the following simple
values~:
\begin{itemize}
 \item[1)] $p\simeq 1\quad q\sim s\sim -a$
 \item[2)]  $p\simeq 1\quad q\sim s\sim 1/a\;$.
\end{itemize}
This leads to definite predictions. Case 2) is especially interesting
because we could show qualitatively in paper I that large $m_t$
induces large neutrino mixing. This is seen explicitly also in the
detailed numerical calculations. The results for
$\nu_e$-$\nu_\mu$ mixing in the cases 1) and 2) for an intermediate
scale of $\sim 10^{12}$ GeV are given in fig.~\ref{fig : mix_12_a}
and fig.~\ref{fig : mix_12_b_old}. They are shown to lie in the
parameter range needed for the MSW explanation~\cite{msw} of the
solar neutrino problem. Fig.~\ref{fig : mix_12_b_new} shows that
in case 2) we can get nice MSW solutions for an intermediate
scale of $\sim 10^{10}$ GeV as well. In fig.~\ref{fig : mix_12_a} and
fig.~\ref{fig : mix_12_b_new} we see that our solutions correspond
exactly to the two regions\footnote{Also fig.~\ref{fig : mix_12_b_old}
corresponds actually to a possible small region in Fig.~1 of the
GALLEX paper. In Fig.~\ref{fig : mix_23_b}, however, it can be seen
that then the depletion of atmospheric neutrinos measured
in the Kamiokande experiment cannot be explained simultaneously (shift
all solutions $\sim 3$ orders of magnitude down to small values in
$\Delta m^2$).} allowed by the recently published results of the
GALLEX collaboration~\cite{GALLEX}~:
\be
 (83\pm 19\;{\rm stat.}\;\pm 8\;{\rm syst.})\; {\rm SNU}.
\ee
Now we are in the position to determine all solutions of
our model compatible with the new
GALLEX data. For the cases 1) and 2) the $\nu_\mu$-$\nu_\tau$ mixing
of these solutions is shown in fig.~\ref{fig : mix_23_a} and
fig.~\ref{fig : mix_23_b} together with the experimental bounds of
Kamiokande and Frejus~\cite{kam}. We find that only in case 2) the
depletion of solar neutrinos as well as atmospheric neutrinos can be
explained simultaneously. A detailed study of the solutions which obey
all those requirements leads to a value of the intermediate scale
between $10^{10}$ GeV and $10^{11}$ GeV.

\clearpage

{\bf Figures}\\
\begin{figure}[h]
 \vspace{8.3cm}
 \caption{$\nu_e$-$\nu_\mu$ mixing in the case 1) for an intermediate
          scale $\sim 10^{12}$ GeV. The curves describe iso--SNU
          lines for $^{71}$Ga detectors.}
 \label{fig : mix_12_a}
\end{figure}
\begin{figure}[h]
 \vspace{8.7cm}
 \caption{$\nu_e$-$\nu_\mu$ mixing in the case 2) for an intermediate
          scale $\sim 10^{12}$ GeV. The curves describe iso--SNU
          lines for $^{71}$Ga detectors.}
 \label{fig : mix_12_b_old}
\end{figure}
\begin{figure}[h]
 \vspace{8cm}
 \caption{$\nu_e$-$\nu_\mu$ mixing in the case 2) for an intermediate
          scale $\sim 10^{10}$ GeV. The curves describe iso--SNU
          lines for $^{71}$Ga detectors.}
 \label{fig : mix_12_b_new}
\end{figure}
\begin{figure}[h]
 \vspace{9.5cm}
 \caption{$\nu_\mu$-$\nu_\tau$ mixing in the case 1) for all solutions
          compatible with GALLEX data and experimental bounds by
          Kamiokande and Frejus.}
 \label{fig : mix_23_a}
\end{figure}
\begin{figure}[h]
 \vspace{8.5cm}
 \caption{$\nu_\mu$-$\nu_\tau$ mixing in the case 2) for all solutions
          compatible with GALLEX data and experimental bounds by
          Kamiokande and Frejus.}
 \label{fig : mix_23_b}
\end{figure}

\end{document}